\documentclass[numreferences]{kluwer}

\usepackage{amsfonts,graphicx}

\input{tcilatex}

\begin{document}

\begin{opening}         
\title{Thermodynamic Bethe ansatz and form factors for the
homogeneous sine-Gordon models \thanks{Talk held at the NATO Advanced Research 
Workshop ``Integrable Structures of Exactly Solvable Two-Dimensional Models of 
Quantum Field Theory", in Kiev (Ukraine), September 25-30 (2000) }} 
\author{Andreas \surname{Fring}}  
\runningauthor{Andreas Fring}
\runningtitle{Thermodynamic Bethe Ansatz and Form Factors for the
HSG Models}
\institute{Institut f\"ur Theoretische Physik, Freie Universit\"at Berlin,
Arnimalle 14, \\ D-14195 Berlin}
\date{January, 2001}

\begin{abstract}
We provide a brief characterization of the main features
of the homogeneous sine-Gordon models and discuss a
general construction principle for colour valued S-matrices,
associated to a pair of simply laced Lie algebras, which contain
the homogeneous sine-Gordon models as a subclass. We give
a brief introduction to the thermodynamic Bethe ansatz and
the form factor approach and discuss explicit solutions
for both methods related to the homogeneous Sine-Gordon
models and its generalization.      
\end{abstract}

\end{opening}           

\section{Introduction}

The work I want to report about \cite{CFKM,FK,CFK,CF1,CF2,CF3} is based on a
collaboration with Olalla Castro-Alvaredo (Universidad de Santiago de
Compostela), Christian Korff (Stony Brook) and Luis Miramontes (Universidad
de Santiago de Compostela).

The completion of the entire bootstrap program \cite{boot} for integrable
quantum field theories in 1+1 space-time dimensions remains an open
challenge for most concrete models. Roughly speaking ``the program'' can be
divided into five distinct steps: i) the determination of the exact on-shell
S-matrix, ii) the computation of closed formulae for the $n$-particle form
factors, iii) the identification of the entire local operator content, iv)
the computation of the related correlation functions and v) various
non-perturbative consistency checks, like the thermodynamic Bethe ansatz,
which exploit the information provided by the related conformal field
theories. Step v) may of course also be carried out by resorting to standard
old fashioned perturbative computations for various quantities. However,
this is slightly opposed to the main virtue and appeal of this approach,
which is its entirely non-perturbative nature in the coupling constant.

Recently we investigated \cite{CFKM,FK,CFK,CF1,CF2,CF3} a class of models,
the homogeneous Sine-Gordon model (HSG) \cite{HSG} and its generalization,
for which the bootstrap program was completed to a large extend. In
comparison with other models, describable by a well-defined Lagrangian and
which have been studied so far in 1+1 dimensions, the HSG-models possess two
interesting novel features familiar from realistic 1+3 dimensional theories.
Namely, their spectrum contains unstable particles and parity invariance is
broken. With regard to one of the motivations for the study of lower
dimensional theories, namely to treat them as test laboratories for general
concepts and ideas, these properties deserve further consideration. One
should say that unstable particles have first been investigated in 1+1
dimensions in the context of the so-called roaming (staircase) models \cite
{Stair} and parity breaking was previously observed for the Federbush models 
\cite{Fed}, in which the scattering matrix is however just a trivial
constant, i.e. rapidity independent, phase.

Due to the limited amount of time and space I will focus in my account on
the main findings, referring the reader to the original literature \cite
{CFKM,FK,CFK,CF1,CF2,CF3} and \cite{PhDChris,LuisProc,PhDOla} for more
details and general background.

\section{The homogeneous sine-Gordon models}

I want to start by recalling the main features of the $G_{k}$-HSG-models 
\cite{HSG} related to simply laced Lie algebras. The latter have been
constructed as integrable perturbations of gauged WZNW-coset theories \cite
{Witten} of the form $G_{k}/U(1)^{\ell }$, where $G$ is a compact simple Lie
group of rank $\ell $ and $k>1$ the level of the Kac-Moody algebra. The
defining action of the HSG-models reads 
\begin{equation}
S_{\text{HSG}}[g]=S_{\text{WZNW}}[g]+\frac{m^{2}}{\pi \beta ^{2}}\,\int
d^{2}x\,\,\left\langle \Lambda _{+},g(\vec{x})^{-1}\Lambda _{-}g(\vec{x}%
)\right\rangle \ \,\;.  \label{HSGaction}
\end{equation}
Here $S_{\text{WZNW}}$ denotes the coset action, $\left\langle
\,\,,\,\,\right\rangle $ the Killing form of $G$ and $g(\vec{x})$ a group
valued bosonic scalar field. $\Lambda _{\pm }$ are arbitrary semi-simple
elements of the Cartan subalgebra associated with the maximal abelian torus $%
H\subset G$, which have to be chosen not orthogonal to any root of $G$. The
parameters $m$ and $\beta ^{2}=1/k+O(1/k^{2})$ are the bare mass scale and
the coupling constant, respectively.

The study of this theory from a classical point of view has been carried out
already to a very large extend and constitutes meanwhile a subject in its
own right. Here I just want to focus on the properties which makes it an
interesting theory from the quantum field theoretical point of view. In
fact, as will be argued in the next sections, from that standpoint, one
could abandon the classical picture altogether, since it just provides
complementary but not necessary information. This is of course the natural
algebraic quantum field theorist's viewpoint.

The classical equations of motion of these models correspond to non-abelian
affine Toda equations \cite{HSG,Nonab}, which are known to be classically
integrable and admit soliton solutions. The integrability on the quantum
level was established in \cite{HSG2} by the construction of non-trivial
conserved charges.

The coset conformal field theory defined by $S_{\text{WZNW}}[g]$ is
characterized by data which are in principle extractable from the
thermodynamic Bethe ansatz (TBA) and the form factor analysis, namely the
Virasoro central charge $c$ of the coset and the conformal dimensions $%
\Delta $ of various primary fields \cite{Witten} 
\begin{equation}
c_{G_{k}/U(1)^{\ell }}=\frac{k-1}{k+h}h\ell ,\,\quad \quad \Delta (\Lambda
,\lambda )=\frac{(\Lambda \cdot (\Lambda +2\rho ))}{2(k+h)}-\frac{(\lambda
\cdot \lambda )}{2k}\;.  \label{cdata}
\end{equation}
Here $h$ is the Coxeter number, $\rho $ the Weyl vector, i.e. the sum over
all fundamental weights, $\Lambda $ a highest dominant weight of level
smaller or equal to $k$ and $\lambda $ their corresponding lower weights
obtained in the usual way by subtracting multiples of simple roots $\alpha
_{i}$ from $\Lambda $ until the lowest weight is reached. The specific
choice of the groups ensures that these theories possess a mass gap \cite
{HSG2}. Of particular interest is the dimension of the perturbing operator,
that is the second term in (\ref{HSGaction}), which is associated to the
conformal dimension related to the adjoint representation $\Delta =\psi
\equiv $ the longest root of $G\,$. We obtained the unique dimension $\Delta
(\psi ,0)=h/(k+h)$. {\small \ }Since $\Delta (\psi ,0)<1$ for all allowed
values of $k$, this perturbation is always relevant in the sense of
renormalisation. To compute the conformal dimensions $\Delta (\Lambda
,\lambda )$ of the entire primary field content is in general quite a
formidable task. We report here only the $SU(3)_{2}/U(1)^{2}$ example with $%
c=6/5$, which we treat in detail in several different approaches. We
obtained \cite{CF1} 12 fields with dimension 1/10, 6 fields of dimension
1/2, 0 and one field, i.e. the perturbing operator of dimension $\Delta
(\psi ,0)=3/5$.

Having established that the theory defined by (\ref{HSGaction}) constitutes
an integrable massive quantum field theory suggests the existence of a
factorizable scattering matrices which can be associated to it. Indeed,
based on the assumption that the semi-classical spectrum is exact, the
S-matrix elements have been conjectured in \cite{HSGS} and verified by means
of the bootstrap principle for HSG-models related to simply laced Lie
algebras. Instead of just stating these results, I would like to provide a
slightly broader context and present a recently found type of scattering
matrix of which the one of \cite{HSGS} is a special case.

\section{Colour valued S-matrices}

For this purpose I want to recall briefly the key features of step i) of the
bootstrap approach. The fundamental observation, on which all further
analysis relies upon, is that integrability, that means here the existence,
one does not need to know its explicit form, of at least one non-trivial
conserved charge, in 1+1 space-time dimensions implies the factorization of
the n-particle scattering matrix into a product of two particle scattering
matrices 
\begin{equation}
Z_{\mu _{n}}(\theta _{n})\ldots Z_{\mu _{1}}(\theta _{1})\left|
0\right\rangle _{\text{out}}=\!\!\!\!\!\!\prod\limits_{1\leq i<j\leq
n}\!\!\!\!\!\!S_{\mu _{i}\mu _{j}}(\theta _{ij})Z_{\mu _{1}}(\theta
_{1})\ldots Z_{\mu _{n}}(\theta _{n})\left| 0\right\rangle _{\text{in}}\,.
\label{fact}
\end{equation}
It is usually convenient to parameterize the two-momentum $\vec{p}$ by the
rapidity $\theta $ as $\vec{p}=m(\cosh \theta ,\sinh \theta )$. We
abbreviate $\theta _{AB}:=\theta _{A}-\theta _{B}$. The $Z_{\mu }(\theta )$
are creation operators for stable particles of type $\mu $ with rapidity $%
\theta $, which obey the Zamolodchikov algebra. The basic assumption of the
bootstrap program is now that every solution to the unitary-analyticity,
crossing and fusing bootstrap equations\footnote{%
For the purpose of this talk I suppose that there is no backscattering in
the theory such that the Yang-Baxter equation constitutes no constraint.} 
\begin{equation}
S_{AB}(\theta )=S_{BA}(-\theta )^{-1}=S_{B\bar{A}}(i\pi -\theta ),\quad
\prod_{l=A,B,C}S_{Dl}(\theta +i\pi \eta _{l})=1\,\,\,,  \label{boot}
\end{equation}
($\eta _{l}\in \Bbb{Q}$ are the fusing angles encoding the mass spectrum and
the anti-particle of $A$ is $\bar{A}$), which admits a consistent
explanation of all poles inside the physical sheet (that is $0<\func{Im}%
\theta <\pi $), leads to a local quantum field theory. There exists no
rigorous proof for this assumption, however, it is supported by numerous
explicitly constructed examples. We may now exploit this principle and
construct new solutions \cite{FK}. For this purpose we supply each particle
type with an additional substructure and identify each particle by two
quantum numbers, i.e. $A\equiv (a,i)$, such that the scattering matrices are
of the form $S_{ab}^{ij}(\theta )$. We associate the main quantum numbers $%
a,b$ to the vertices of the Dynkin diagram of a simply laced Lie algebra 
\textbf{g }of rank $\ell $ and the colour quantum numbers $i,j$ to the
vertices of the Dynkin diagram of a simply laced Lie algebra \textbf{\~{g}}
of rank $\tilde{\ell}$. Noting now that for many (not always though)
theories the scattering matrix factorizes into a so-called minimal part $%
S_{AB}^{\min }(\theta )$, which satisfies (\ref{boot}) by itself, and into a
CDD-factor $S_{AB}^{\text{CDD}}(\theta ,B(\beta ))$ which depends on the
effective coupling constant $B(\beta )$ containing only unphysical poles, $%
S_{AB}(\theta )=S_{AB}^{\min }(\theta )S_{AB}^{\text{CDD}}(\theta ,B(\beta
)) $, we may include the colour structure by defining the interaction as 
\begin{equation}
S_{ab}^{ij}(\theta )=\QATOPD\{ . {S_{ab}^{\min }(\theta
)=(S_{ab}^{CDD}(\theta ,B_{ii}=0))^{-1}\qquad \text{for }i=j}{S_{ab}^{CDD}(%
\theta ,B_{ij})\qquad \qquad \qquad \,\qquad \qquad \text{for }i\neq
j}\,\,\,.  \label{5}
\end{equation}
Allowing now \textbf{g,\~{g} }to be simply laced algebras, we found \cite{FK}
as solution 
\begin{equation}
S_{ab}^{ij}(\theta )=e^{i\pi \varepsilon _{ij}K_{\bar{a}b}^{-1}+\int%
\nolimits_{-\infty }^{\infty }\frac{dt}{t}\left( 2\cosh \frac{\pi t}{h}-%
\tilde{I}\right) _{ij}\left( 2\cosh \frac{\pi t}{h}-I\right)
_{ab}^{-1}e^{-it(\theta +\sigma _{ij})}}.
\end{equation}

\noindent The theory possess various free parameters, the $\ell $ mass
scales of the stable particles and the $\tilde{\ell}$ mass scales of the
unstable particles characterized by the resonance parameters $\sigma
_{ij}=-\sigma _{ji}$ ($1\leq i,j\leq \tilde{\ell}$) with the additional
constraint that $i$ and $j$ are connected on the \textbf{\~{g}}-Dynkin
diagram.

As special cases of these models we have the \textbf{g\TEXTsymbol{\vert}}$%
\mathbf{A}_{1}$-theories and the $\mathbf{A}_{n}$\textbf{\TEXTsymbol{\vert}%
\~{g}}-theories, which coincide with the minimal affine Toda theories and
the \textbf{\~{g}}$_{n+1}$-HSG-models, respectively. S-matrices which allow
the ``colour algebra'' \textbf{\~{g} }also to be non-simply laced were
recently proposed in \cite{CK}. The generalization which also admits \textbf{%
g} to be non-simply laced still remains an interesting open problem. In that
case the outlined factorization into minimal and CDD-factor does not take
place anymore and one therefore has to abandon the described construction
principle.

As already mentioned as particular example, which we want analyze in more
detail, we consider the $\mathbf{A}_{1}$\textbf{\TEXTsymbol{\vert}}$\mathbf{A%
}_{2}$ $\equiv $ $SU(3)_{2}$-HSG-model with 
\begin{equation}
S_{\pm \pm }=-1\quad \quad \text{,}\quad \quad S_{\pm \mp }(\theta )=\pm
\tanh \left( \theta \pm \sigma -i\pi /2\right) /2\;.
\end{equation}
Since there is no pole in the physical sheet, fusing processes do not take
place. An important point to note here, to which we frequently appeal, is
that for the limit $\sigma \rightarrow \infty $ the S-matrices S$_{\pm \mp }$
tend to one and the whole theory decouples into a direct product of two
thermally perturbed Ising models.

\section{Description of unstable particles}

Since the states in (\ref{fact}) are asymptotic, it is clear that unstable
particles have to be described by different means, even when they possess a
very long life time. As is familiar from standard quantum mechanics, one may
introduce a decay width by complexifying the mass of a particle. As is for
instance explained in \cite{ELOP}, the same prescription can be taken over
to quantum field theory. The description of an unstable particle of type $%
\tilde{c}$ may be thought of as adding a decay width $\Gamma _{\tilde{c}}$
to the physical mass of a stable particle, such that the S-matrix as a
function of the Mandelstam $s$-variable has a pole at $s=M_{R}^{2}=(M_{%
\tilde{c}}{}-i\Gamma _{\tilde{c}}/2)^{2}$. Transforming from the $s$ to the
rapidity plane and describing the scattering of two stable particles of type 
$a\,$and $b$ with masses $m_{a}$ and $m_{b}$ by $S_{ab}(\theta )$, the
resonance pole is situated at $\theta _{R}=\sigma -i\bar{\sigma}$.
Identifying the real and imaginary parts of the pole then yields 
\begin{eqnarray}
M_{\tilde{c}}^{2}{}-\Gamma _{\tilde{c}}^{2}/4
&=&m_{a}^{2}{}+m_{b}^{2}{}+2m_{a}m_{b}\cosh \sigma \cos \bar{\sigma},
\label{BW1} \\
M_{\tilde{c}}\Gamma _{\tilde{c}} &=&2m_{a}m_{b}\sinh |\sigma |\sin \bar{%
\sigma}\,\,.  \label{BW2}
\end{eqnarray}
Whenever $M_{\tilde{c}}\gg \Gamma _{\tilde{c}}$, the quantity $M_{\tilde{c}}$
admits a clear cut interpretation as the physical mass and the relations (%
\ref{BW1}) and (\ref{BW2}) acquire the form which is usually referred to as
Breit-Wigner formula \cite{BW}. Since this assumption is only required for
interpretational reasons we will not demand it in general. Eliminating the
decay width from (\ref{BW1}) and (\ref{BW2}), we can express the mass of the
unstable particles $M_{\tilde{c}}$ in the model as a function of the masses
of the stable particles $m_{a},m_{b}$ and the resonance parameter $\sigma $.
Assuming $\sigma $ to be large we obtain 
\begin{equation}
M_{\tilde{c}}^{2}\sim m_{a}m_{b}(1+\cos \bar{\sigma})/2\,\,e^{|\sigma |}\,.
\label{Munst}
\end{equation}

The occurrence of the variable $me^{|\sigma |/2}$ is interesting, since it
was introduced originally in \cite{triZam} in order to describe massless
particles, i.e. one may perform safely the simultaneous limit $m\rightarrow
0,\sigma \rightarrow \infty $, and one might therefore be tempted to
describe flows related to (\ref{Munst}) as so-called massless flows. In
fact, in \cite{CFKM} we introduced the variable explicitly as a formal
parameter.

One should note, however, a few essential differences between the roaming or
staircase models \cite{Stair} and the HSG models. First of all the decay
width of the unstable particles enters in a different manner. Whereas in the
HSG-model they are introduced by a rapidity shift $\theta \rightarrow \theta
\pm \sigma $ of a parity broken theory, in the staircase models they enter
through an analytic continuation at the self-dual point of the effective
coupling constant $B$, i.e. $B\rightarrow 1\pm i\sigma /\pi $, of scattering
matrices associated to affine Toda field theories. Concerning the analytic
continuation, the point $B=1$ is quite special, since for this point the
combination $B(B-2)$ remains real. The other distinction between the two
classes of models is the origin of the staircase pattern observed in the
scaling functions of the models (see below). For the HSG-models one may
associate the steps directly to the energy scale of the unstable particles,
which is not possible for the staircase models. Furthermore, the HSG-models
admit a well-defined Lagrangian. The condition $\Gamma _{\tilde{c}}\ll M_{%
\tilde{c}}$ corresponds to the semi-classical limit.

\section{Thermodynamic Bethe Ansatz}

Originally the TBA was formulated in the context of the non-relativistic
Bose gas by Yang and Yang \cite{Yang} in order to extract various
thermodynamic quantities. Thereafter it was extended \cite{TBAZam1} to
relativistic quantum field theories whose scattering matrices factorize into
two-particle ones. In this context the approach allows the construction of a
scaling function which reproduces some characteristic features, to be
specified in more detail below, of the model and in particular various
quantities of the underlying ultraviolet conformal field theory, in
particular the effective Virasoro central charge and sometimes also the
dimension of the perturbing operator.

It is well-known that in a relativistically invariant theory one can in
general not use a wavefunction formalism, because of the creation of real
and virtual particles. However, in configuration space one may assume to
have regions in which one regards the particles as free and describe them
formally by a function $\psi (x_{1},\ldots ,x_{N})$, $N$ being the total
number of particles. Two adjacent free regions are then connected by an
S-matrix. Taking now the particle $A$ on a trip around the world, a circle
in this case, the formal wavefunction of $A$ picks up the corresponding
S-matrix element as a phase factor when meeting another particle. For the
case at hand we have to distinguish, whether the particle is moved clockwise
or counter-clockwise along the world line, due to the breaking of parity
invariance. Proceeding this way, we obtain two types of Bethe Ansatz
equations 
\begin{equation}
e^{iLM_{A}\sinh \theta _{A}}\prod\limits_{B\neq A}S_{AB}(\theta
_{AB})=e^{-iLM_{A}\sinh \theta _{A}}\prod\limits_{B\neq A}S_{BA}(\theta
_{BA})=1\,,  \label{BA}
\end{equation}
with $L$ denoting the length of the compactified space direction. These two
sets of equations are of course not entirely independent and may be obtained
from each other by complex conjugation and noting that $S$ is
Hermitian-analytic \cite{David}. Unlike as for instance in the talk of
Slavnov (see this proceeding), we do not use the equations (\ref{BA}) to
find allowed momenta, but instead we take the logarithm of these equations
and carry out the thermodynamic limit in the usual fashion \cite{TBAZam1},
i.e. letting the size of the system go to infinity $L\rightarrow \infty $,
while keeping the ratio $L/N$ finite. This way we obtain the following sets
of coupled non-linear integral equations 
\begin{equation}
\sum\nolimits_{B}\,\Phi _{AB}\ast L_{B}^{\pm }(\theta ,r)=r\,M_{A}\cosh
\theta +\ln (1-\exp (-L_{A}^{\pm }(\theta ,r)))\quad \,.  \label{tba}
\end{equation}
Here the rapidity convolution of two functions is denoted by $f\ast g(\theta
):=\int d\theta ^{\prime }/2\pi \,f(\theta -\theta ^{\prime })g(\theta
^{\prime })$. We also re-defined the masses by $M_{A}\rightarrow M_{A}/m_{1}$
keeping the same notation. The parameter $r=m_{1}T^{-1}$ is the inverse
temperature times the overall mass scale of the lightest particle. The
kernels in the integrals carry the information of the dynamical and
statistical interaction of the system and are given by 
\begin{equation}
\Phi _{AB}(\theta )=\Phi _{BA}(-\theta )\;=-i\frac{d\ln S_{AB}(\theta )}{%
d\theta }\,-2\pi g_{AB}\delta (\theta ).  \label{kernel}
\end{equation}
For generic values of $g_{AB}$ the statistical interaction is of Haldane
type \cite{BF}. The special cases $g_{AB}=0$ and $g_{AB}=\delta _{AB}$
correspond to bosonic and fermionic statistics, respectively. Recently also
a formulation for Gentile statistics has been proposed \cite{Andrei}. Here
we will restrict our attention to the choice $g_{AB}=\delta _{AB}$. As very
common in these considerations, we could also parameterize the function $%
L_{A}^{\pm }(\theta )=\ln (1+e^{-\epsilon _{A}^{\pm }(\theta )})$ by the
so-called pseudo-energies $\epsilon _{A}^{\pm }(\theta )$. Then it follows
from the properties of the TBA equations that $\epsilon _{A}^{+}(\theta
)=\epsilon _{A}^{-}(-\theta )$. In comparison with the parity invariant case
the main difference is that we have lost the usual symmetry of the
pseudo-energies as a function of the rapidities, since we have $\epsilon
_{A}^{+}(\theta )\neq \epsilon _{A}^{-}(\theta )$ in this case. The main
task of the TBA is now to solve the equations (\ref{tba}) for the $L$'s.
Thereafter, one may compute the scaling function 
\begin{equation}
c(r)=\frac{3\,r}{\pi ^{2}}\sum_{A}M_{A}\int\nolimits_{0}^{\infty }d\theta
\,\cosh \theta \,(L_{A}^{-}(\theta )+L_{A}^{+}(\theta ))\,,  \label{scale}
\end{equation}
whose key property is 
\begin{equation}
\lim_{r\rightarrow 0}c(r)=c_{\text{eff}}=c-24\Delta ^{\prime }\,,
\end{equation}
with $c$ being the Virasoro central charge and $\Delta ^{\prime }$ the
lowest occurring conformal dimension of the underlying ultraviolet conformal
field theory. We recall that for unitary models the values of $c$ and $c_{%
\text{eff}}$ coincide.

Due to the non-linear nature of the equations (\ref{tba}), there exist
hardly any exact analytical solutions for the full TBA-equations, albeit it
is possible to find various analytical approximations, e.g. \cite{APP}. When
one is only interested in $c_{\text{eff}}$, it is possible to solve (\ref
{tba}) and (\ref{scale}) fully analytically. Noting that the TBA-equations
for the \textbf{g\TEXTsymbol{\vert}\~{g}}-theories develop constant regions
in this limit one may write (\ref{tba}) in this case in the very symmetric
form 
\begin{equation}
\prod_{b=1}^{\ell }\left( Q_{b}^{i}\right) ^{I_{ab}^{g}}+\prod_{j=1}^{\tilde{%
\ell}}\left( Q_{a}^{j}\right) ^{I_{ij}^{\tilde{g}}}=\left( Q_{a}^{i}\right)
^{2}\,\,.  \label{QQ}
\end{equation}
Here we introduced the variable $Q_{a}^{i}=\prod_{b=1}^{\ell
}(L_{b}^{i}{})^{(K_{ab}^{\tilde{g}})^{-1}}$. For finite values of the
resonance parameter the effective central charge is then expressible as 
\begin{equation}
c_{\text{eff}}^{\mathbf{g}|\mathbf{\tilde{g}}} =\frac{\ell \tilde{\ell}\,%
\tilde{h}}{h+\tilde{h}} 
=\frac{6}{\pi ^{2}}\sum\limits_{a=1}^{\ell }\sum\limits_{i=1}^{\tilde{\ell}%
}\mathcal{L}\left( 1-\prod\nolimits_{b=1}^{\ell }\left( Q_{b}^{i}\right)
^{-K_{ab}^{g}}\right) \,\,  \label{cc2}
\end{equation}
with $\mathcal{L}(x)=\sum_{n=1}^{\infty }x^{n}/n^{2}+\ln x\ln (1-x)/2$
denoting Rogers dilogarithm. Remarkably, besides the numerical solutions,
for many combinations of \textbf{g }and \textbf{\~{g}}, the relations (\ref
{QQ}) have been solved analytically \cite{Resh,Kun,CFKM,CF6} in such a way
that the $Q_{a}^{i}$'s acquire the form of certain Weyl-characters.
Recalling that the $\mathbf{A}%
_{n}$\textbf{\TEXTsymbol{\vert}\~{g}}-theories$\equiv $ \textbf{\~{g}}$_{n+1}
$-HSG-models we note that the first part of(\ref{cdata}) and (\ref{cc2}) 
obviously coincide.

From a purely mathematical point of view the system of equations (\ref{QQ})
and second part of (\ref{cc2}) is an exceptional situation 
whenever $c_{\text{eff}}$ is
rational, in which case it is referred to as accessible dilogarithms (for a
review see e.g. \cite{dilog} and references therein). Remarkably, the same
system of equations may be obtained by employing techniques originally
pursuit in \cite{Rich}.

\begin{center}
\includegraphics[width=11.5cm,height=9.14cm]{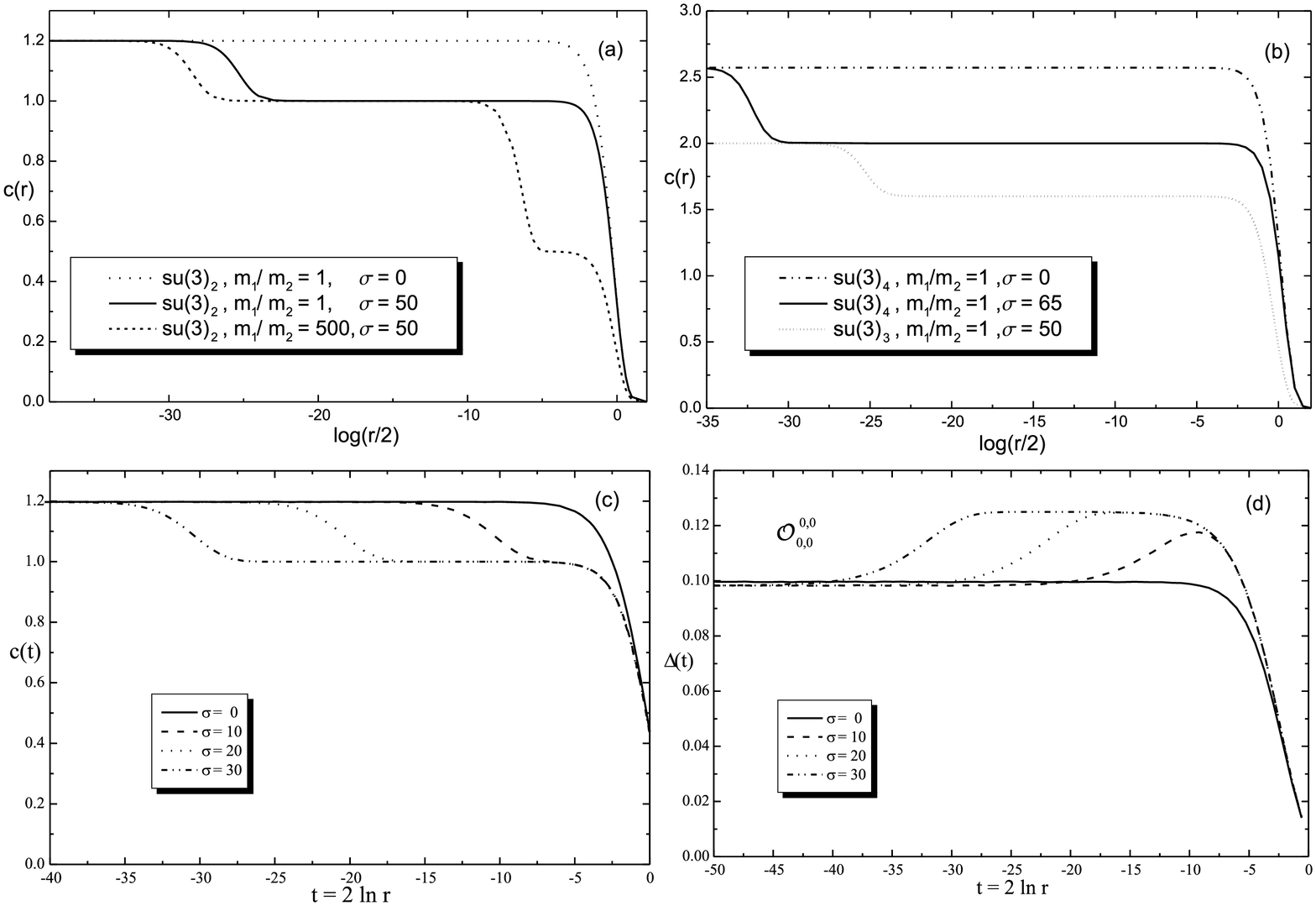}
\end{center}

\noindent {\tiny Figure 1: Scaling functions for $su(3)_{k},\;k=2,3,4$ from
the thermodynamic Bethe ansatz as a function of the variable $\log r/2$ with 
$r=m_{1}T^{-1}$ at different values of the resonance shift and mass ratios
(a), (b). Scaling function from the correlation function $\left\langle
\Theta (r)\Theta (0)\right\rangle $ for $su(3)_{2}$ (c) and RG-flow of the
conformal dimension of the local operator $\mathcal{O}_{0,0}^{0,0}$ (d) as a
function of 2 $\log r$.}

In \cite{CFKM} we analyzed the full TBA-equations for the \textbf{A}$_{k-1}$%
\textbf{\TEXTsymbol{\vert}A}$_{2}$-theory$\equiv $ SU(3)$_{k}$-HSG. The
scaling function may be approximated in this case by 
\begin{equation}
c(r,M_{\tilde{c}}^{\tilde{k}})\approx \left\{ 
\begin{array}{ll}
4\,\frac{k-1}{k+2}\,\,, & \qquad \text{for\quad }1\ll \frac{2}{r}\ll M_{%
\tilde{c}}^{\tilde{k}} \\ 
6\,\frac{k-1}{k+3}\,, & \qquad \text{for\quad }M_{\tilde{c}}^{\tilde{k}}\ll 
\frac{2}{r}
\end{array}
\right. \,,\,  \label{platt}
\end{equation}
where $M_{\tilde{c}}^{\tilde{k}}$ is the smallest mass of an unstable bound
state which may be formed in the process $(a,i)+(b,j)\rightarrow (\tilde{c},%
\tilde{k})$. This analytical approximation is confirmed by our full
numerical solution as depicted in figure 1 (a), (b).

For the different algebras treated, we note that the different plateaus in
the scaling functions are characterized by the mass scales of the unstable
particles as described in (\ref{platt}). By choosing the masses of the two
particles to be very different we also confirm in part (a) the decoupling of
the theory into a direct product of two thermally perturbed Ising models.
Exploiting the observation \cite{CFKM,CF2,CF3} that the plateaus are also
realizable as cosets, it is also possible to find analytical descriptions
similar in spirit to (\ref{QQ}) and (\ref{cc2}) for them \cite{CF6}.

\section{Form Factors}

To use form factors as auxiliary objects in order to compute correlation
functions is an approach originally pioneered by the Berlin group in the
late seventies \cite{Kar}. Form factors are tensor valued functions,
representing matrix elements of some local operator $\mathcal{O}(\vec{x})$
at the origin between a multiparticle in-state and the vacuum, which we
denote by

\begin{equation}
F_{n}^{\mathcal{O}|\mu _{1}\ldots \mu _{n}}(\theta _{1},\ldots ,\theta
_{n}):=\left\langle 0|\mathcal{O}(0)|Z_{\mu _{1}}(\theta _{1})Z_{\mu
_{2}}(\theta _{2})\ldots Z_{\mu _{n}}(\theta _{n})\right\rangle _{\text{in}%
}\,\,.
\end{equation}
Here the $Z_{\mu }(\theta )$'s are the same type of creation operators as in
Eq. (\ref{fact}).

One way to determine the form factors, that is step ii) in the bootstrap
program, is similar in spirit to the computation of the S-matrix, namely one
considers the consequences various physical concepts have on these matrix
elements. This way one can set up a system of constraints, which, like in
the case of the scattering matrix, turns out to be so restrictive that they
lead to their explicit determination. We shall now only recall these
constraints and refer for a more systematic and rigorous derivation of them
to \cite{BFKZ}.

As a consequence of CPT-invariance or the exchange of two operators $Z_{\mu
}(\theta )$ one obtains 
\begin{equation}
F_{n}^{\mathcal{O}|\ldots \mu _{i}\mu _{j}\ldots }(\ldots \theta _{i},\theta
_{j}\ldots )=F_{n}^{\mathcal{O}|\ldots \mu _{j}\mu _{i}\ldots }(\ldots
\theta _{j},\theta _{i}\ldots )S_{\mu _{i}\mu _{j}}(\theta _{i,j})\,.
\label{W1}
\end{equation}
The analytic continuation in the complex $\theta $-plane at the threshold
cuts when $\theta =2\pi i$ together with crossing leads to 
\begin{equation}
F_{n}^{\mathcal{O}|\mu _{1}\ldots \mu _{n}}(\theta _{1}+2\pi i,\ldots
,\theta _{n})=F_{n}^{\mathcal{O}|\mu _{2}\ldots \mu _{n}\mu _{1}}(\theta
_{2},\ldots ,\theta _{n},\theta _{1})\,\,.  \label{W2}
\end{equation}
Since we are describing relativistically invariant theories, we expect for
an operator $\mathcal{O}$ with spin $s$%
\begin{equation}
F_{n}^{\mathcal{O}|\mu _{1}\ldots \mu _{n}}(\theta _{1}+\lambda ,\ldots
,\theta _{n}+\lambda )=e^{s\lambda }F_{n}^{\mathcal{O}|\mu _{1}\ldots \mu
_{n}}(\theta _{1},\ldots ,\theta _{n})\,\,,  \label{rel}
\end{equation}
with $\lambda $ being an arbitrary real number. For a form factor whose
first two particles are conjugate to each other we have the so-called
kinematical pole at $i\pi $, which leads to a recursive equation relating
the (n-2)- and the n-particle form factor 
\begin{eqnarray}
&&-i\stackunder{{\small \bar{\theta}}_{0}\rightarrow {\small \theta }_{0}}{%
\text{Res}}{\small F}_{n+2}^{\mathcal{O}|\bar{\mu}\mu \mu _{1}\ldots \mu
_{n}}{\small (\bar{\theta}}_{0}{\small +i\pi ,\theta }_{0}{\small ,\theta }%
_{1}{\small ,\ldots ,\theta }_{n}{\small )}=  \nonumber \\
&&\qquad \qquad \qquad (1-\omega \prod\nolimits_{l=1}^{n}S_{\mu \mu
_{l}}(\theta _{0l}))\,{\small F}_{n}^{\mathcal{O}|\mu _{1}\ldots \mu _{n}}%
{\small (\theta }_{1}{\small ,\ldots ,\theta }_{n}{\small ),}  \label{kin}
\end{eqnarray}
with $\omega $ being the factor of local commutativity and $\bar{\mu}$ the
anti-particle of $\mu $. We restrict our initial considerations to a model
in which stable bound states may not be formed and therefore do not need to
report on the so-called bound state residue equation, which relates an
(n-1)- and an n-particle form factor.

The equations (\ref{W1})-(\ref{kin}) allow already the determination of
several solutions. However, since these equations are completely independent
of $\mathcal{O}(\vec{x})$, one needs further arguments to establish which
type of operators one deals with. This means we are in a quite different
position in comparison with the computation of matrix elements involving
some well-known operator and Bethe states as described for instance in the
talk by Kitanine (see this proceeding). There are however additional
properties, which do involve the nature of the operator. In general they
make use of their asymptotic behaviour, like the momentum space cluster
property. It states that whenever the first, say $\kappa $, rapidities of an 
$n$-particle form factor are shifted to infinity, the $n$-particle form
factor factorizes into a $\kappa $ and an ($n-\kappa $)-particle form
factor, which are possibly related to different types of local operators 
\begin{equation}
\mathcal{T}_{1,\kappa }^{\lambda }F_{n}^{\mathcal{O}}(\theta _{1},\ldots
,\theta _{n})\sim F_{\kappa }^{\mathcal{O}^{\prime }}(\theta _{1},\ldots
,\theta _{\kappa })F_{n-\kappa }^{\mathcal{O}^{\prime \prime }}(\theta
_{\kappa +1},\ldots ,\theta _{n})\,\,.  \label{cluster}
\end{equation}
We have introduced here the operator $\mathcal{T}_{a,b}^{\lambda
}:=\lim_{\lambda \rightarrow \infty }\prod_{p=a}^{b}T_{p}^{\lambda }$ $\,$%
which is composed of the translation operator $T_{a}^{\lambda }$ acting on a
function of $n$ variables as $T_{a}^{\lambda }\,f(\theta _{1},\ldots ,\theta
_{a},\ldots ,\theta _{n})\,\mapsto \,f(\theta _{1},\ldots ,\theta
_{a}+\lambda ,\ldots ,\theta _{n})$. For the purely bosonic case this
behaviour can be explained perturbatively by means of Weinberg's power
counting theorem, see e.g. \cite{Kar,BK}. For a similar restricted
situation, there exists also a short heuristic argument which provides some
form of intuitive picture of this behaviour \cite{DSC} by appealing to the
ultraviolet conformal theory. However, this argument is based on various so
far unjustified assumptions, which need further clarification. In addition,
up to now it has never been made use of. It is very interesting that when (%
\ref{cluster}) is read from the r.h.s. to the l.h.s. it may been seen a kind
of fusion ($\neq $ the fusing mentioned in section 3 which refers to the
particles) of the local operators. Therefore, in order to shed more light on
step iii) of the bootstrap program, it would be important to put the
property (\ref{cluster}) on the same solid ground as the equations (\ref{W1}%
)-(\ref{kin}).

Having determined the form factors one may carry out step iv) of the
bootstrap program and expand the two-point correlation function as
\smallskip\ 
\begin{eqnarray}
&&\left\langle \mathcal{O}(r)\mathcal{O}^{\prime }(0)\right\rangle
=\sum_{n=1}^{\infty }\sum_{\mu _{1}\ldots \mu _{n}}\int\nolimits_{-\infty
}^{\infty }\frac{d\theta _{1}\ldots d\theta _{n}}{n!(2\pi )^{n}}e^{-r\,E} 
\nonumber \\
&&\times \,\,F_{n}^{\mathcal{O}|\mu _{1}\ldots \mu _{n}}(\theta _{1},\ldots
,\theta _{n})\,\left( F_{n}^{\mathcal{O}^{\prime }|\mu _{1}\ldots \mu
_{n}}(\theta _{1},\ldots ,\theta _{n})\,\right) ^{\ast }  \label{corr}
\end{eqnarray}
where $E=\sum\nolimits_{i=1}^{n}m_{\mu _{i}}\cosh \theta _{i}$ is the sum of
the on-shell energies. Provided we can compute (\ref{corr}) we may use the
two-point correlation function in order to determine the RG-flow of the
Virasoro central charge $c$ and the conformal dimension of a primary field.
According to the c-theorem of Zamolodchikov we have 
\begin{eqnarray}
c(r) &=&\frac{3}{2}\int\nolimits_{r}^{\infty }ds\,s^{3}\,\,\left\langle
\Theta (s)\Theta (0)\right\rangle  \\
&=&3\sum_{n=1}^{\infty }\sum_{\mu _{1}\ldots \mu _{n}}\int\nolimits_{-\infty
}^{\infty }\frac{d\theta _{1}\ldots d\theta _{n}}{n!(2\pi )^{n}}e^{-r\,E} 
\nonumber \\
&&\times \left| F_{n}^{\Theta |\mu _{1}\ldots \mu _{n}}(\theta _{1},\ldots
,\theta _{n})\right| ^{2}\frac{(6+6rE+3r^{2}E^{2}+r^{3}E^{3})}{2E^{4}}
\label{dellc}
\end{eqnarray}
where $\Theta $ is the trace of the energy momentum tensor. The flow of a
conformal weight of a primary field\smallskip \smallskip\ was proposed in 
\cite{CF2} 
\begin{eqnarray}
\Delta (r) &=&-\frac{1}{2\left\langle \mathcal{O}(0)\right\rangle }%
\int\nolimits_{r}^{\infty }ds\,s\,\,\left\langle \Theta (s)\mathcal{O}%
(0)\right\rangle \,\,  \\
&=&-\sum_{n=1}^{\infty }\sum_{\mu _{1}\ldots \mu _{n}}\int\nolimits_{-\infty
}^{\infty }\frac{d\theta _{1}\ldots d\theta _{n}}{n!(2\pi )^{n}}\frac{%
(1+rE)e^{-r\,E}}{2\left\langle \mathcal{O}(0)\right\rangle E^{2}}  \nonumber
\\
&&\,\,\,\times F_{n}^{\Theta |\mu _{1}\ldots \mu _{n}}(\theta _{1},\ldots
,\theta _{n})\left( F_{n}^{\mathcal{O}|\mu _{1}\ldots \mu _{n}}(\theta
_{1},\ldots ,\theta _{n})\,\right) ^{\ast }\,  \label{delta}
\end{eqnarray}
as a straightforward generalization of the $\Delta $-sum rule \cite{DSC}.
The conformal dimension of a primary field may also be obtained by a direct
analysis of the well-known ultraviolet behaviour of the correlation function 
\begin{equation}
\lim_{r\rightarrow 0}\left\langle \mathcal{O}(r)\mathcal{O}(0)\right\rangle
\sim r^{-4\Delta ^{\!\!\mathcal{O}}}\,\,.  \label{ult}
\end{equation}
The main difference between the two possibilities is that (\ref{ult}) leads
to far less conclusive values as (\ref{delta}). Whereas (\ref{ult}) requires
in general an a priori knowledge of $\Delta ^{\!\!\mathcal{O}}$, and serves
mainly to confirm this, the expression (\ref{delta}) can just be evaluated by
brute force.

I will now present the explicit solutions found for equations (\ref{W1})-(%
\ref{ult}) for the $SU(3)_{2}$-HSG model \cite{CFK,CF1,CF2}. A step towards
a more group theoretical understanding was done in \cite{CF3}, where these
results were generalized to the $SU(N)_{2}$ case.

Labeling an operator by four quantum numbers $\mu ,\nu ,\tau ,\tau ^{\prime
}$ we proved in \cite{CF2} that the general n-particle solution reads 
\begin{eqnarray}
&&F_{2s+\tau ,2t+\tau ^{\prime }}^{\mathcal{O}_{\tau ,\tau ^{\prime }}^{\mu
,\nu }|M^{+}M^{-}}(\theta _{1},\ldots ,\theta _{n})=H_{2s+\tau ,2t+\tau
^{\prime }}^{\mathcal{O}_{\tau ,\tau ^{\prime }}^{\mu ,\nu
}|M^{+}M^{-}}\,\!\!\det \mathcal{A}_{2s+\tau ,2t+\tau ^{\prime }}^{\mu ,\nu }
\nonumber \\
&&\left( \sigma _{2s+\tau }^{+}\right) ^{s-t+\frac{\tau -1-\nu }{2}}\left(
\sigma _{2t+\tau ^{\prime }}^{-}\right) ^{\frac{1+\tau -\tau ^{\prime }-\mu 
}{2}-t}\prod_{i<j}\hat{F}^{\mu _{i}\mu _{j}}(\theta _{ij})\,.  \label{solu}
\end{eqnarray}
W.l.g. we assumed here a particular ordering by starting with $2s+\tau $
particles of the type $\mu =+$ followed by $2s+\tau ^{\prime }$ particles of
the type $\mu =-$, collected in the sets $M^{\pm }=\{\pm ,\ldots ,\pm \}$.
Once these expressions are known, all other form factors related to it by
permutations of the particles may be constructed trivially by exploiting (%
\ref{W1})-(\ref{W2}). The functions $\hat{F}^{\mu _{i}\mu _{j}}$ for all
combinations of the $\mu $'s are 
\begin{eqnarray}
\hat{F}^{\pm \pm }(\theta ) &=&-i/2\tanh \frac{\theta }{2}\exp (\mp \theta
/2) \\
\hat{F}^{\pm \mp }(\theta ) &=&2^{\frac{1}{4}}e^{\tfrac{i\pi (1\mp 1)\pm
\theta }{4}-\tfrac{G}{\pi }-\int\nolimits_{0}^{\infty }\tfrac{dt}{t}\tfrac{%
\sin ^{2}\left( (i\pi -\theta \mp \sigma )\frac{t}{2\pi }\right) }{\sinh
t\cosh t/2}},  \label{14}
\end{eqnarray}

\noindent with $G$ being the Catalan constant. The ($t+s$)$\times $($t+s$%
)-matrix 
\begin{equation}
\left( \mathcal{A}_{2s+\tau ,2t+\tau ^{\prime }}^{\mu ,\nu }\right)
_{ij}=\QATOPD\{ . {\sigma _{2(j-i)+\mu }^{+}\text{,\quad }1\leq i\leq t}{%
\hat{\sigma}_{2(j-i)+2t+\nu }^{-}\,\,\,\text{,\quad }t<i\leq s+t}
\label{Acomp}
\end{equation}
has as its entries elementary symmetric polynomials depending on different
sets of variables. We use the notation $\sigma ^{\pm }$ when they depend on
the variable $x=\exp \theta $ associated to the sets $M^{\pm }$ and $\hat{%
\sigma}$ to indicate that all variables are multiplied by a factor $%
ie^{-\sigma }$. The overall constant was computed to 
\begin{equation}
H_{2s+\tau ,2t+\tau ^{\prime }}^{\mathcal{O}_{\tau ,\tau ^{\prime }}^{\mu
,\nu }}=i^{s(2\tau +\tau ^{\prime }+\nu +2)}2^{s(2s-2t-\tau ^{\prime
}-1+2\tau )}e^{s\sigma (2t+\tau ^{\prime })/2}H_{\tau ,2t+\tau ^{\prime }}^{%
\mathcal{O}_{\tau ,\tau ^{\prime }}^{\mu ,\nu }}\,\,\,\,
\end{equation}
\noindent where the value of $H_{\tau ,2t+\tau ^{\prime }}^{\mathcal{O}%
_{\tau ,\tau ^{\prime }}^{\mu ,\nu }}$ is fixed by the lowest non-vanishing
form factor. To determine this constant requires additional arguments. That
these expressions are indeed solutions of (\ref{W1})-(\ref{kin}) was
rigorously proven in \cite{CF2}. Alternatively we showed that the following
cluster properties hold 
\begin{eqnarray}
\mathcal{T}_{1,2\kappa +\xi \leq l}^{\lambda }F_{2s+\tau ,2t+\tau ^{\prime
}}^{\mu ,0} &\sim &F_{2\kappa +\xi ,0}^{0,0}F_{2s+\tau -2\kappa -\xi
,2t+\tau ^{\prime }}^{\mu +\xi (1-2\mu ),0},  \label{40} \\
\mathcal{T}_{1,2\kappa +\xi \leq l}^{-\lambda }F_{2s+\tau ,2t+\nu }^{\mu
,\nu } &\sim &F_{2\kappa +\xi ,0}^{0,0}F_{2s+\tau -2\kappa -\xi ,2t+\nu
}^{\mu ,\nu }, \\
\mathcal{T}_{n+1-2\kappa -\xi <m,n}^{\lambda }F_{2s+\tau ,2t+\tau ^{\prime
}}^{0,\nu } &\sim &F_{2s+\tau ,2t+\tau ^{\prime }-2\kappa -\xi }^{0,\nu +\xi
(1-2\nu )}F_{0,2\kappa +\xi }^{0,0}, \\
\mathcal{T}_{n+1-2\kappa -\xi <m,n}^{-\lambda }F_{2s+\mu ,2t+\tau ^{\prime
}}^{\mu ,\nu } &\sim &F_{2s+\mu ,2t+\tau ^{\prime }-2\kappa -\xi }^{\mu ,\nu
}F_{0,2\kappa +\xi }^{0,0}\,,  \label{41}
\end{eqnarray}
where we simplified the notation in an obvious way. The equations 
(\ref{40})-(\ref{41}) may be seen on one hand as a consistency check, since
we recover the solutions constructed directly, and also as a construction
principle, because from one particular solution we may derive various other
ones, i.e. we simulate the ``fusion'' process.

Since the form factors of the trace of the energy-momentum tensor occur in
various applications it is instructive to use it as an example to give the
idea how more explicit formulae look like when all the quantities are
assembled together 
\begin{equation}
F_{2s,2t}^{\Theta }=\sigma _{1}(x_{1},\ldots ,x_{n})\sigma
_{1}(x_{1}^{-1},\ldots ,x_{n}^{-1})F_{2s,2t}^{\mathcal{O}_{2,2}^{1,1}}\,\,.
\end{equation}
To fix the initial condition is in general also still a difficult issue, but
we may solve this problem here easily by appealing to the fact that when the
n-particle form factors involve only one type of particle they have to
coincide with the ones for the thermally perturbed Ising model. The only
non-vanishing form factor for the trace of the energy-momentum tensor is 
\begin{equation}
F_{2}^{\Theta }(\theta )=-2\pi im^{2}\sinh (\theta /2)\,\,\,\,.
\end{equation}
Explicit expressions for  4,6-particle form factors are for instance 
\begin{eqnarray}
F_{4}^{\Theta |++--} &=&\tfrac{-\pi m_{-}^{2}e^{(\theta _{31}+\theta
_{42})/2}(2+\sum_{i<j}\cosh (\theta _{ij}))}{2\cosh (\theta _{12}/2)\cosh
(\theta _{34}/2)}\prod_{i<j}F_{\text{min}}^{\mu _{i}\mu _{j}}(\theta _{ij})\,
\\
F_{6}^{\Theta |++++--} &=&\frac{\pi m^{2}(3+\sum_{i<j}\cosh (\theta _{ij}))}{%
4\prod_{1\leq i<j\leq 4}\cosh (\theta _{ij}/2)}\prod_{i<j}\tilde{F}_{\text{%
min}}^{\mu _{i}\mu _{j}}(\theta _{ij})\,.
\end{eqnarray}

\noindent The matrix $\mathcal{A}$ is in this case of dimension ($t+s-2$)$%
\times $($t+s-2$) and reads explicitly 
\begin{equation}
\mathcal{A}^{\Theta }=\left( 
\begin{array}{rrrrrr}
\sigma _{1}^{+} & \sigma _{3}^{+} & \sigma _{5}^{+} & \sigma _{7}^{+} & 
\cdots & 0 \\ 
0 & \sigma _{1}^{+} & \sigma _{3}^{+} & \sigma _{5}^{+} & \cdots & 0 \\ 
\vdots & \vdots & \vdots & \vdots & \ddots & \vdots \\ 
0 & 0 & 0 & 0 & \cdots & \sigma _{2s-1}^{+} \\ 
-\hat{\sigma}_{1}^{-} & \hat{\sigma}_{3}^{-} & -\hat{\sigma}_{5}^{-} & \hat{%
\sigma}_{7}^{-} & \cdots & 0 \\ 
0 & -\hat{\sigma}_{1}^{-} & \hat{\sigma}_{3}^{-} & -\hat{\sigma}_{5}^{-} & 
\cdots & 0 \\ 
\vdots & \vdots & \vdots & \vdots & \ddots & \vdots \\ 
0 & 0 & 0 & 0 & \cdots & (-1)^{t}\hat{\sigma}_{2t-1}^{-}
\end{array}
\right) \,\,\,.
\end{equation}
Exploiting the integral representation for the symmetric polynomials $\sigma
_{k}(x_{1},\ldots ,x_{n})\,=\frac{1}{2\pi i}\oint_{|z|=\varrho }\frac{dz}{%
z^{n-k+1}}\prod\nolimits_{k=1}^{n}(z+x_{k})$, it is easy to express the
determinant of $\mathcal{A}^{\Theta }$ as a multidimensional integral
representation 
\begin{eqnarray}
&&\!\!\!\!\det \mathcal{A}^{\Theta }\mathcal{\,\,}=(-1)^{(s+1)t}\oint
du_{1}\ldots \oint du_{t-1}\oint dv_{1}\ldots \oint dv_{s-1}\,\,\,  \nonumber
\\
&&\times
\prod\limits_{j=1}^{t-1}u_{j}^{2-2s-2j}\prod\limits_{i=1}^{2s}(u_{j}+x_{i})%
\prod\limits_{j=1}^{s-1}v_{j}^{2-2t-2j}\!\!\!\!\prod%
\limits_{i=1+2s}^{2s+2t}(v_{j}+\hat{x}_{i})  \nonumber \\
&&\times \!\!\!\!\prod_{1\leq i<j\leq
t-1}(u_{j}^{2}-u_{i}^{2})\!\!\!\!\prod_{1\leq i<j\leq
s-1}(v_{j}^{2}-v_{i}^{2})\prod_{j=1}^{s-1}%
\prod_{i=1}^{t-1}(u_{i}^{2}+v_{j}^{2})\,\quad \,\,
\end{eqnarray}
$\medskip $where we abbreviated $\oint dz\equiv (2\pi
i)^{-1}\oint\nolimits_{|z|=\varrho }dz$ with $\varrho $ being a positive
real number. These type of integrals differ from the ones provided in \cite
{BFKZ}, where the integration is carried out in the $\theta $ rather than
the $x$-variables. To manifest the precise relation between the two
formulations remains an open issue at present.

Indicating by a superscript the number of the highest n-particle form
factors taken into account in (\ref{dellc}) and (\ref{delta}), we computed
by means of a Monte Carlo integration

\begin{eqnarray}
c(0)^{(2)} &=&1,\,\,\,c(0)^{(4)}=1.197...,\,\,\,c(0)^{(6)}=1.199\ldots
\,,\quad \text{ }\sigma <\infty \,\,\quad \\
\lim_{\sigma \rightarrow \infty }c(0) &=&1\,.
\end{eqnarray}
The latter equation is exact, since we have $\lim_{\sigma \rightarrow \infty
}F_{2s+2t}^{\Theta |2s,2t}\sim e^{-(t+s-1)\sigma }$, from which we deduce
that only $F_{2}^{\Theta |0,2}$ and $F_{2}^{\Theta |2,0}$ are non-vanishing,
which allows an analytic evaluation. We obtained \cite{CF1} 
\begin{equation}
\Delta ^{\!\!\mathcal{O}_{0,0}^{0,0}}(0)^{(6)}=0.1004\qquad \quad \text{ }%
\sigma <\infty \,\ .
\end{equation}
The entire RG-flow of $c$ and $\Delta ^{\!\!\mathcal{O}_{0,0}^{0,0}}$ is
depicted in figure 1. We observe two crucial features: First of all, by
comparing part (a) and (c) we see that the c-flow determined from (\ref
{dellc}) is qualitatively identical to the one obtained from the TBA.
Secondly by comparing part (c) and (d) we observe that whenever the $%
SU(3)_{2}$-HSG model decouples into a direct product of two thermally
perturbed Ising models, the $\mathcal{O}_{0,0}^{0,0}$-field decouples into
two fields with conformal dimension 1/16, i.e. the disorder operator of the
Ising model. We confirm these findings alternatively by a direct analysis of
(\ref{ult}). A similar computation \cite{CF1} was carried out also for
several other solution (\ref{solu}). Unfortunately, we did not find
solutions corresponding to all primary fields, in particular we did not find
any solution corresponding to $\Delta =1/2$.

\section{Conclusions and open issues}

The main conclusion which can be drawn from our analysis concerning the
specific status of the HSG-models (\textbf{g\TEXTsymbol{\vert}\~{g}}%
-theories) is that the scattering matrix originally proposed in \cite{HSGS}
is certainly consistent and can be associated to a perturbed gauged
WZNW-coset. This is based on the fact that we reproduce all the predicted
features of this picture, namely the expected ultraviolet Virasoro central
charge, various conformal dimensions of local operators and the
characteristics of the unstable particle spectrum by means of the TBA as
well as the form factor analysis.

The key statement concerning a comparison of the two different methods is
that they yield the same qualitative behaviour of the scaling function. With
regard to the efficiency of the methods one should point out that in the TBA
approach the number of coupled non-linear integral equations to be solved
increases with the number of different particles, which means the system
becomes extremely complex and cumbersome to solve even numerically.
Computing the scaling function with the help of form factors only adds more
terms to each $n$-particle contribution, but is technically not more
involved. The price we pay in this setting is, however, a slower convergence
of (\ref{corr}). With regard to other quantities, that is the flow of the
conformal dimensions, one should note that in the TBA-approach it is not
known how to solve this problem and even at the fixed point itself one is,
so far, only able to determine the conformal dimension of the perturbing
operator with considerable effort for some particular theories. In the
context of form factors there is no obstacle to the computation, once the
solution is known.

Concerning the status of the bootstrap program, one should say that a very
rich structure can be associated to step i). In particular, there exist
closed general group theoretical expressions, which allow a relatively
universal understanding as opposed to model dependent statements.
Nonetheless. a complete classification is still outstanding. The status of
step ii) is still less satisfactory since computable closed formulae exist
only in few cases and one is still far from a closed group theoretical
formulation. Step iii) also still requires some considerations, since up to
now it is not even possible to identify properly the operators related to
primary fields (we did not find various dimensions and can not undo the
degeneracy), not to mention all the remaining operators. Further study of
the cluster property will certainly provide advance in this direction. The
computation of correlation functions, i.e. step iv) can in principle be
carried out numerically, but of course a better analytical understanding is
highly desirable. Our results \cite{CF3} also indicate that the
``folkloristic belief'' of the fast convergence of the series expansion of (%
\ref{corr}) has to be challenged. In fact, for large values of $N$, this is
not true anymore. It would be highly desirable to have more concrete
quantitative criteria at hand. Needless to say that it would be nice to
increase amount of technical tools available to carry out v) and enrich the
spectrum of non-perturbative arguments.

\end{document}